\documentstyle[12pt]{article}

\catcode`\@=11
\long\def\@makefntext#1{
\protect\noindent \hbox to 3.2pt {\hskip-.9pt
$^{{\ninerm\@thefnmark}}$\hfil}#1\hfill}		

 \def\@makefnmark{\hbox to 0pt{$^{\@thefnmark}$\hss}}  

\def\ps@myheadings{\let\@mkboth\@gobbletwo
\def\@oddhead{\hbox{}
\rightmark\hfil\ninerm\thepage}
\def\@oddfoot{}\def\@evenhead{\ninerm\thepage\hfil
\leftmark\hbox{}}\def\@evenfoot{}
\def\sectionmark##1{}\def\subsectionmark##1{}}

\newcounter{sectionc}\newcounter{subsectionc}\newcounter{subsubsectionc}
\renewcommand{\section}[1] {\vspace{0.6cm}\addtocounter{sectionc}{1}
\setcounter{subsectionc}{0}\setcounter{subsubsectionc}{0}\noindent
	{\bf\thesectionc. #1}\par\vspace{0.4cm}}
\renewcommand{\subsection}[1] {\vspace{0.6cm}\addtocounter{subsectionc}{1}
	\setcounter{subsubsectionc}{0}\noindent
	{\it\thesectionc.\thesubsectionc. #1}\par\vspace{0.4cm}}
\renewcommand{\subsubsection}[1]
{\vspace{0.6cm}\addtocounter{subsubsectionc}{1}
	\noindent {\rm\thesectionc.\thesubsectionc.\thesubsubsectionc.
	#1}\par\vspace{0.4cm}}

\newcounter{appendixc}
\newcounter{subappendixc}[appendixc]
\newcounter{subsubappendixc}[subappendixc]

\renewcommand{\appendix}[1] {\vspace{0.6cm}
        \refstepcounter{appendixc}
        \setcounter{figure}{0}
        \setcounter{table}{0}
        \setcounter{equation}{0}
        \renewcommand{\thefigure}{\Alph{appendixc}.\arabic{figure}}
        \renewcommand{\thetable}{\Alph{appendixc}.\arabic{table}}
        \renewcommand{\theappendixc}{\Alph{appendixc}}
        \renewcommand{\theequation}{\Alph{appendixc}.\arabic{equation}}
        \noindent{\bf Appendix \theappendixc #1}\par\vspace{0.4cm}}

\def\abstracts#1{{
	\centering{\begin{minipage}{30pc}\tenrm\baselineskip=12pt\noindent
	\centerline{\tenrm ABSTRACT}\vspace{0.3cm}
	\parindent=0pt #1
	\end{minipage}}\par}}

\renewenvironment{thebibliography}[1]
	{\begin{list}{\arabic{enumi}.}
	{\usecounter{enumi}\setlength{\parsep}{0pt}
\setlength{\leftmargin 1.25cm}{\rightmargin 0pt}
	 \setlength{\itemsep}{0pt} \settowidth
	{\labelwidth}{#1.}\sloppy}}{\end{list}}

\newcounter{itemlistc}
\newcounter{romanlistc}
\newcounter{alphlistc}
\newcounter{arabiclistc}

\newcommand{\fcaption}[1]{
        \refstepcounter{figure}
        \setbox\@tempboxa = \hbox{\tenrm Fig.~\thefigure. #1}
        \ifdim \wd\@tempboxa > 6in
           {\begin{center}
        \parbox{6in}{\tenrm\baselineskip=12pt Fig.~\thefigure. #1}
            \end{center}}
        \else
             {\begin{center}
             {\tenrm Fig.~\thefigure. #1}
              \end{center}}
        \fi}

\newcommand{\tcaption}[1]{
        \refstepcounter{table}
        \setbox\@tempboxa = \hbox{\tenrm Table~\thetable. #1}
        \ifdim \wd\@tempboxa > 6in
           {\begin{center}
        \parbox{6in}{\tenrm\baselineskip=12pt Table~\thetable. #1}
            \end{center}}
        \else
             {\begin{center}
             {\tenrm Table~\thetable. #1}
              \end{center}}
        \fi}

\def\@citex[#1]#2{\if@filesw\immediate\write\@auxout
	{\string\citation{#2}}\fi
\def\@citea{}\@cite{\@for\@citeb:=#2\do
	{\@citea\def\@citea{,}\@ifundefined
	{b@\@citeb}{{\bf ?}\@warning
	{Citation `\@citeb' on page \thepage \space undefined}}
	{\csname b@\@citeb\endcsname}}}{#1}}

\newif\if@cghi
\def\cite{\@cghitrue\@ifnextchar [{\@tempswatrue
	\@citex}{\@tempswafalse\@citex[]}}
\def\citelow{\@cghifalse\@ifnextchar [{\@tempswatrue
	\@citex}{\@tempswafalse\@citex[]}}
\def\@cite#1#2{{$\null^{#1}$\if@tempswa\typeout
	{IJCGA warning: optional citation argument
	ignored: `#2'} \fi}}

\def\fnt#1#2{\footnotetext{\kern-.3em
	{$^{\mbox{\sevenrm #1}}$}{#2}}}

 1
 1
 1

\font\tenbf=cmbx10
\font\tenrm=cmr10
\font\tenit=cmti10

\font\ninerm=cmr9

\begin{document}
\pagestyle{empty}

\hspace{3cm}

\vspace{-3.4cm}
\rightline{{ FTUAM 95/5}}

\vspace{1.0cm}

\centerline{\tenbf EXTRACTING PREDICTIONS FROM SUPERGRAVITY/SUPERSTRING}
\baselineskip=16pt
\centerline{\tenbf FOR THE EFFECTIVE THEORY BELOW THE PLANCK SCALE
\footnote{Plenary talk given at the International Conference "Beyond
the Standard Model IV", Lake Tahoe (California), December 13--18, 1994.}}
\vspace{0.8cm}
\centerline{\tenrm C. MU\~NOZ}
\baselineskip=13pt
\centerline{\tenit Departamento de F\'{\i}sica Te\'{o}rica C-XI, Universidad
Aut\'{o}noma de Madrid, Cantoblanco}
\baselineskip=12pt
\centerline{\tenit Madrid, E-28049, Spain}
\vspace{0.9cm}
\abstracts{The connection between Superstring theory and the low--energy
world is analyzed. In particular, the soft Supersymmetry--breaking terms
arising in Supergravity theories coming from Superstrings are computed.
Several solutions proposed to solve the $\mu$ problem,
and the $B$ soft term associated, are discussed. The issue of gauge
coupling constants unification in the context of Superstrings is also
discussed.}

\vspace{1.5cm}
\begin{flushleft}
{FTUAM 95/5} \\
{March 1995}
\end{flushleft}

\vfil
\rm\baselineskip=14pt
\section{Introduction and Summary}

The most outstanding virtue of Superstring theory is that
it is the only (finite) theory which can unify all the known
interactions including gravity.
Furthermore, it is the only hope to answer fundamental questions that in the
context of the Standard Model (SM) or Grand Unified Theories (GUTs)
cannot even be posed:
why the gauge and Yukawa couplings should have a particular value?. First,
the gauge coupling constant is dynamical because it arises as the
vacuum expectation value (VEV) of
a gauge singlet field $S$ called the dilaton, $\langle Re S\rangle = 1/g^2$.
Second, the Yukawa couplings, which determine the
quark and lepton masses, can be explicitly calculated and
they turn out to be
also dynamical. They depend in general on other gauge singlet fields $T_m$
called the moduli whose VEV
determine the size and shape of the
compactified space. E.g. for the overall
modulus $\langle Re T\rangle \sim R^2$. Of course,
experimental data demand
$Re S \sim 2$ and $Re T \sim 1$ (in Planck mass units). Therefore the initial
questions translate as how are the VEV determined. This will be discussed
below. Finally, it is possible to obtain models resembling the Supersymmetric
Standard Model (SSM) at
low--energy. Of
course, this is crucial in order to connect Superstring theory with the
observable
world.

The particle spectrum in the SSM is in general determined by the
soft Supersymmetry (SUSY)--breaking terms. In the simplest SSM, the
so--called Minimal Supersymmetric Standard Model (MSSM), assuming
certain universality of soft terms these can be
parametrized by only
four parameters: a universal gaugino mass $M$, a universal scalar mass $m$, a
universal trilinear
scalar parameter $A$ and an extra bilinear scalar parameter $B$.
These soft terms are very important not only because they determine the SUSY
spectrum, like gaugino, squark and
slepton masses, but also because they contribute to the
Higgs potential generating the
radiative breakdown of the electroweak symmetry.

When Supergravity (SUGRA) is spontaneously broken in a "hidden sector"
the soft SUSY-breaking terms are generated. These are
characterized by the gravitino mass ($m_{3/2}$) scale and therefore in order
not to introduce a problem of naturalness, $m_{3/2}$ should be of the
electroweak scale order (recall that the soft terms contribute to the
Higgs masses). An interesting non--perturbative source of
SUSY--breaking, capable of generating
this large mass hierarchy, is gaugino condensation in some hidden sector
gauge group.

The full SUGRA Lagrangian is specified in terms of two functions which depend
on the hidden and observable scalars of the theory:
the real analytic gauge--invariant K\"ahler function $G$ which is a combination
of two functions $K$ (the K\"ahler potential) and
$W$ (the superpotential), and the analytic gauge
kinetic function $f$.
Then, once we know these functions the soft SUSY--breaking terms are
calculable.
For example, for the simple case of
canonical kinetic terms for hidden and observable fields and constant
$f$, it is straightforward to compute the form of the soft terms\cite{Hall}
\begin{eqnarray}
M	&=& 0 \nonumber\\
m^2	&=& m_{3/2}^2 + V_0 \nonumber\\
A       &=& m_{3/2} \sum_l h_l^* {G}_{h_l} \nonumber\\
B       &=& A - m_{3/2} \label{F14}
\end{eqnarray}
where $h_l$ are the hidden sector fields,
$V_0$ is the VEV of the scalar potential (i.e. the cosmological
constant), and we use standard SUGRA conventions on
derivatives (e.g. $G_\alpha = \frac{\partial G}{\partial z_\alpha^{*}}$,
$G^\alpha = \frac{\partial G}{\partial z_\alpha}$).
$M=0$ is a consecuence of the assumption that $f$ is a constant.

Unfortunately for the predictivity of the theory, $G$ and $f$ are
arbitrary and therefore the soft terms become SUGRA model dependent.
Besides, the existence of the "hidden sector" has to be postulated "ad hoc".
All the these problems can be solved in Superstring theory.

In particular, the Heterotic Superstring
after compactification of six extra dimensions (on some compact manifolds)
leads to a $N=1$ effective SUGRA. Now, $K$ and $f$ {\it are in principle
calculable}
from Superstring scattering amplitudes. Besides, whereas in SUGRA
(non--Superstring) models we do not have the slightest idea
of what fields could be involved in SUSY--breaking, four--dimensional
{\it Superstring theory automatically has natural candidates for that job}: the
dilaton $S$ and the moduli $T_m$. These gauge singlet fields are
generically present in four--dimensional Heterotic Superstring since $S$
is related with the gravitational sector of the theory and $T_m$ are related
with the extra dimensions. While other extra fields could also play
a role in specific models, the dilaton and moduli constitute in some way
the {\it minimal possible SUSY--breaking sector in Superstring theory}\footnote
{Starting with this minimal sector one can also study the possible role
on SUSY--breaking of other extra fields (see the example discussed in
section 8 of ref.\cite{BIM}).}.

Concerning the superpotential $W$, the situation
is more involved. It is known that the process of SUSY--breaking in
Superstring theory has to have a non--perturbative origin
since SUSY is preserved order by order in perturbation
theory (the scalar potential $V(S,T_m)$ is flat) and hence $S$ and $T_m$ are
undetermined at this level.
On the other hand, very little is known about non--perturbative
effects in Superstring theory, particularly in the four--dimensional case. It
is true
that gaugino condensation gives rise to an effective $W$ which breaks SUSY
at the same time that $S$ and $T_m$ acquire reasonable VEVs\cite{Beatriz}
(as the ones
explained above), and determines
explicitly the values of the soft SUSY--breaking terms\cite{Beatriz2}. However
one should
keep in mind that this analysis requires the assumption that the dominant
non--perturbative effects in Superstring theory are the field theory ones. This
is
because gaugino condensation is not a pure "stringy" mechanism.
Thus a pessimist would say that Superstring theory does not look particularly
promising in trying to get information about the SUSY--breaking sector
of the theory.

However, since $K$ and $f$ are known, and the degrees of
freedom involved in the process of SUSY--breaking have been identified
(the "hidden fields" $S$ and $T_m$), the effect of SUSY--breaking can be
parametrized by the VEVs of those fields. In the next section we
review this approach for addressing the problem, trying to provide
a theory of soft terms which could enable us to interpret the (future)
experimental results on SUSY spectra. It turns out to be specially useful
to introduce several "goldstino angles" whose values tell us where the
dominant source of SUSY--breaking resides. All formulae for soft parameters
take on particularly simple forms when written in terms of these variables.
In section 3 a different kind of connection between Superstring theory and
the low--energy phenomenology is analyzed, namely the possibility of obtaining
the gauge coupling constants unification in Superstring models as the
experimental (LEP) results demand.

\section{Soft terms from Superstring theory}
\vspace{-0.7cm}
\subsection{General structure of soft terms}
$K$ (to first order in the observable fields $\phi_i$)
is given in general by the form
\begin{eqnarray}
K = -log(S+S^*)\ +\ K_0(T,T^*)  + K_{ij}(T,T^*)\phi_i\phi_j^*
\label{F9}
\end{eqnarray}
where the indices $i$, $j$ label the charged matter
fields\footnote{For phenomenological reasons related to the
absence of flavour changing neutral currents (FCNC) in the effective
low--energy theory (see section 5 of ref.\cite{BIM} for a discussion of
this point) from now on we will assume a diagonal form for the part
of the K\"ahler potential associated with matter fields,
$K_{ij}=K_{i}\delta_j^i$ in eq.(\ref{F9}).}.
Amongst the moduli $T_m$ we concentrate, for the moment, on the overall modulus
$T$ whose classical value gives the size of the manifold. Apart from
simplicity, this modulus is the only one which is always necessarily present
in any $(0,2)$ (but left-right symmetric) 4-D Superstrings.
 We believe that
studying the one modulus case is enough to get a feeling of the
most important physics of soft terms. Anyway, the case with several moduli
will be analyzed in subsection 2.4.
We will disregard for the moment any mixing between the $S$ and $T$ fields
kinetic terms. In fact this is strictly correct in all 4-D Superstrings at
tree level. However, it is known that this type of mixing may arise at one
loop level in some cases. On the other hand, these are loop effects
which should
be small and in fact can be easily incorporated in the analysis in some simple
cases (orbifolds) as mention below.

The tree-level expression for $f$ for any four-dimensional Superstring
is well known,
$f_a=k_aS$, where $k_a$ is the Kac-Moody level of the gauge factor. Normally
(level one case) one takes $k_3=k_2=\frac{3}{5}k_1=1$. Since a possible
$T$ dependence may appear at one--loop (higher--loop corrections are
vanishing), then in general
\begin{eqnarray}
f_a(S,T)\ =\ k_aS + f_a(T) \label{F26}
\end{eqnarray}
where we assume that other possible chiral fields do not contribute to
SUSY--breaking.

Finally, the cosmological constant is
\begin{eqnarray}
V_0\ =\ {G}_S^S |F^S|^2\ +\ {G}_T^T |F^T|^2 \ -\ 3 e^{G} \label{F27}
\end{eqnarray}
Of course,
the first two terms in the right hand side of eq.(\ref{F27}) represent the
contributions of the $S$ and $T$ auxiliary fields,
$F^S=e^{G/2}({G}^S_S)^{-1} G^S$ and
$F^T=e^{G/2}({G}^T_T)^{-1} G^T$.

As we will show below, it is important to know what field, either $S$ or $T$,
plays the predominant role in the process of SUSY-breaking. This
will have relevant consequences in determining
the pattern of soft terms, and therefore the spectrum of physical
particles. That is why
it is very useful to  define an angle $\theta $
in the following way\cite{BIM}  (consistently with eq.(\ref{F27})):
\begin{eqnarray}
({G}^S_S)^{1/2}\ F^S\ &=&\ {\sqrt 3}C m_{3/2}\ e^{i\alpha _S}sin\theta
\nonumber\\
({G}^T_T)^{1/2}\ F^T\ &=&\ {\sqrt 3}C m_{3/2}\ e^{i\alpha _T}cos\theta
\label{F28}
\end{eqnarray}
where $\alpha_S,\alpha_T$ are the phases of $F^S$ and  $F^T$, and the
constant $C$ is defined as follows:
\begin{eqnarray}
C^2\ =\ 1\ +\ {{V_0}\over {3m_{3/2}^2}}
\label{F29}
\end{eqnarray}
If the cosmological constant $V_0$ is assumed to vanish, one has $C=1$, but we
prefer for the moment to leave it undetermined. Of course, the way one deals
with the
cosmological constant problem is important.

Notice that, with the above assumptions, the goldstino field which is swallowed
by the gravitino in the process of SUSY--breaking is proportional
to
\begin{eqnarray}
{\tilde {\eta }}\ =\ sin\theta \ {\tilde S}\ +\ cos\theta \ {\tilde T}
\label{F30}
\end{eqnarray}
where ${\tilde S}$ and ${\tilde T}$ are the canonically normalized
fermionic partners of the
scalar fields $S$ and $T$ (we have reabsorbed here the phases by redefinitions
of the fermions ${\tilde S},{\tilde T}$). Thus the angle defined above
may be appropriately termed {\it goldstino angle} and has a clear
physical interpretation as a mixing angle.

Then it is straightforward to compute the general form of the soft
terms\cite{BIM}
\begin{eqnarray}
M_a &=&  {{C{\sqrt 3}}\over {2 Ref_a}} m_{3/2}
( k_a 2 Re S\ e^{-i\alpha _S}sin\theta \
 + \ f_a^T({G_0}_T^T)^{-1/2}\ e^{-i\alpha _T}cos\theta )
\label{F31}\\
m_i^2 &=& 2m_{3/2}^2\ (C^2-1) \ +\ m_{3/2}^2C^2(1 +N_i(T,T^*)cos^2\theta )
\nonumber\\
N_i(T,T^*) &=& {3 \over {{K_0}_T^T} }
( { { {K_i}_T {K_i}^T }
\over { {K_i}^2 }  }  \ -\
{ {{K_i}^T_T}\over  {K_i}  } )
\ = \
-3{{ (log K_i)^T_T} \over {{K_0}_T^T} }
\label{F32}\\
A_{ijk} &=& -{\sqrt 3}m_{3/2}C(e^{-i\alpha _S}sin\theta \  +
\ e^{-i\alpha _T} \omega_{ijk}(T,T^*) cos\theta)
\nonumber\\
{\omega _{ijk}(T,T^*)} &=& ({K_0}_T^T)^{-1/2}
( {\sum _{l=i,j,k}} {{K_l^T}\over {K_l}}
\ -\ K_0^T\ -\ {{Y_{ijk}^T}\over {Y_{ijk}}} \ )
\label{F33}\\
B_{\mu} &=& m_{3/2}(\ -1\ -C{\sqrt 3}e^{-i\alpha _S}sin\theta
(1-{{\mu ^S}\over {\mu }}(S+S^*))\nonumber\\
&+&C{\sqrt 3}e^{-i\alpha _T}cos\theta
({K_0}_T^T)^{-1/2}( K_0^T +\ {{\mu ^T}\over {\mu }}
-{{{K}_{H_1}^T}\over {{K}_{H_1}}}
-{{{K}_{H_2}^T}\over {{K}_{H_2}}}))
\label{F91}
\end{eqnarray}
where $Y_{ijk}$ are the usual Yukawa couplings.
The above expressions become much simpler in specific four--dimensional
Superstrings and/or in the large--T limit. This is the case for instance of
the formula for
$N_i(T,T^*)$ which looks complicated but it becomes very simple.
$N_i$ is related to the curvature of the K\"ahler manifold parametrized by the
above K\"ahler potential. For manifolds of constant curvature (like in
the orbifold case) the $N_i$ are constants, independent of $T$. More precisely,
they correspond to the modular weights of the charged fields, which are
normally negative integer numbers. In more complicated
four-dimensional Superstrings like those based on Calabi-Yau manifolds, the
$N_i(T,T^*)$ functions are complicated expressions in which world-sheet
instanton effects play an important role. In the case of $(2,2)$ Calabi-Yau
manifolds, for the large $T$ limit it turns out that
$N_i(T,T^*)\rightarrow -1$.

Anyway,
the explicit dependence of the soft masses on the $N_i$
of each particle produces a lack of universality\cite{IL,Beatriz2}. This
may be relevant for the issue of flavour-changing neutral currents (FCNC).
For an extended discussion on this point see section 5 of ref.\cite{BIM}
and refs.\cite{Pokorski,Nir}.

The soft terms obtained in the previous analysis
are in general complex.
Notice that if $S$ and $T$ fields acquire complex vacuum expectation
values, then the phases $\alpha _S,\alpha _T$ associated
with their auxiliary fields can be non-vanishing and
the functions $\omega _{ijk}(T,T^*)$, $f_a (T)$, etc.
can be complex. The analysis of this situation in connection with the
experimental limits on the electric dipole moment of the neutron (EDMN)
can be found in refs.\cite{BIM,Choi}.

Finally,
the above analysis has shown that the different soft SUSY-breaking terms
have all an explicit dependence on $V_0$, i.e. the cosmological constant,
which is contained in $C$.
We have to face this fact and do something
about it\footnote{It is worth noticing that
general properties which are independent
of the value of the cosmological constant can still be
found (see subsection 3.3 of ref.\cite{BIM}).}.
We cannot just simply ignore it, as it is often done, since
the way we deal with the cosmological constant problem has a bearing
on measurable quantities like scalar and gaugino masses.
For an extended discussion on this point see sections 6 and 8 of ref.\cite{BIM}
and refs.\cite{KN,Park,Ferrara}.

The value of the $B$ soft term depends on the solution proposed to generate
a $\mu$ term in the low--energy theory.
In ref.\cite{Casas} was pointed out that the presence of a
non--renormalizable term in the superpotential
\begin{eqnarray}
\lambda W H_1 H_2
\label{F18}
\end{eqnarray}
characterized by the coupling $\lambda$, which
mixes the observable sector with the hidden sector, yields dynamically a
$\mu$ parameter when $W$ acquires a VEV
\begin{eqnarray}
\mu = \lambda W
\label{F19}
\end{eqnarray}
The fact that $\mu$ is of the electroweak scale order is a consequence
of our assumption of a correct SUSY--breaking scale
$ m_{3/2} = e^{K/2} |W| = O(M_W) $.

The superpotential eq.(\ref{F18}) which provides a possible solution to the
$\mu$ problem can be naturally obtained in the context of
Superstring theory.
In ref.\cite{Casas} a realistic example where non--perturbative
SUSY--breaking mechanisms like gaugino--squark condensation induce
superpotentials of the type eq.(\ref{F18}) was given. In ref.\cite{Narain}
the same kind of superpotential was obtained using pure gaugino condensation.
It was used the fact that in some classes of four--dimensional Superstrings
(orbifolds) a possible $H_1H_2$ dependence may appear in $f$ at one--loop.
In both cases $\lambda=\lambda(T)$ in general, so with this solution to the
$\mu$ problem eq.(\ref{F91}) gives\cite{BIMS}
(let us call $B_{\lambda}$ the $B$--term
associated with eqs.(\ref{F18},\ref{F19}))
\begin{eqnarray}
B_{\lambda}\ =m_{3/2}(\ (3C^2-1) + C{\sqrt 3}e^{-i\alpha _T}cos\theta
({K_0}_T^T)^{-1/2}(\ {{\lambda ^T}\over {\lambda }}
-{{{K}_{H_1}^T}\over {{K}_{H_1}}}
-{{{K}_{H_2}^T}\over {{K}_{H_2}}}))
\label{F35}
\end{eqnarray}
The alternative mechanism in which there is an extra term in the K\"ahler
potential
\begin{eqnarray}
\delta K = Z H_1 H_2  +\ h.c.
\label{F99}
\end{eqnarray}
originating a $\mu$--term\cite{Masiero,Casas}
is also naturally present in some large classes of four--dimensional
Superstrings. Then, the $B$--term (we will call it $B_Z$) is given
by\cite{BIM}
\begin{eqnarray}
B_Z &=& \frac{m_{ 3/2}}{X} ((3C^2-1) +
C {\sqrt 3}e^{-i\alpha _T}cos\theta
({K_0}_T^T)^{-1/2}({{Z^T}\over Z}-{{{K}_{H_1}^T}\over
{{K}_{H_1}}}-{{{K}_{H_2}^T}\over {{K}_{H_2}}})\nonumber\\
&-&C{\sqrt 3}e^{i\alpha _T}cos\theta ({K_0}_T^T)^{-1/2}{{Z_T}\over Z}
+C^2 3({K_0}_T^T)^{-1}
cos^2\theta ({{Z_T}\over Z}({{{K}_{H_1}^T}\over
{{K}_{H_1}}}+{{{K}_{H_2}^T}\over {{K}_{H_2}}})\
-\ {{Z^T_T}\over Z}))
\nonumber\\
X &\equiv& 1- C \sqrt{3} e^{i \alpha_T} \cos\theta
({K_0}^T_T)^{-1/2} \frac{Z_T}{Z}
\label{F98}
\end{eqnarray}
Indeed, in the case of some orbifold models and the large--T
limit of Calabi--Yau compactifications one expect\cite{Louis,Dieter,Narain}
\begin{eqnarray}
Z \simeq \frac {1}{T+T^*}
\label{F36}
\end{eqnarray}

Notice that it is conceivable that both mechanisms could be present
simultaneously.
In that case the general expresions for the $B$--term and Higgsino
mass $\widehat{\mu}$ are easily obtained\cite{BIMS}
\begin{eqnarray}
B &=& \frac{1}{\widehat{\mu}}
    ( B_{\lambda} m_{3/2} \lambda + B_Z m_{3/2} XZ ) (K_{H_1} K_{H_2})^{-1/2}
\label{F24}\\
\widehat{\mu} &=& ( m_{3/2} \lambda + m_{3/2} X Z ) (K_{H_1} K_{H_2})^{-1/2}
\label{F24}
\end{eqnarray}
where $B_{\lambda}$ and $B_Z$, $X$ are given in eqs.(\ref{F35}) and (\ref{F98})
respectively.

\subsection{The $sin\theta =1$ (dilaton-dominated) limit}
Before going into specific classes of String models, it is worth studying the
interesting limit $sin\theta =1$, corresponding to the case where
the dilaton sector is the source of all the SUSY-breaking\cite{Louis,BIM} (see
eq.(\ref{F28})). Since the
dilaton couples in a universal manner to all particles, this limit is
quite model independent.
Using eqs.(\ref{F31}--\ref{F91},\ref{F35},\ref{F98})
one finds the following simple expressions for the
soft terms:
\begin{eqnarray}
M_a &=& {\sqrt 3}Cm_{3/2}{{k_a ReS}\over {Ref_a}}e^{-i\alpha _S}
\nonumber\\
m_i^2 &=& C^2 m_{3/2}^2\ +\ 2m_{3/2}^2(C^2-1)
\nonumber\\
A_{ijk} &=& -{\sqrt 3}Cm_{3/2}e^{-i\alpha _S}
\nonumber\\
B_{\mu } &=& m_{ 3/2}( -1\ -\ {\sqrt 3}Ce^{-i\alpha _S}
(1-{{\mu ^S}\over {\mu }}(S+S^*)))
\nonumber\\
B_{\lambda} &=& B_Z =\ m_{ 3/2}( 3C^2-1)
\label{F37}
\end{eqnarray}
Notice that the scalar masses and the $A$--terms are universal,
whereas the gaugino masses may be slightly non-universal
since non-negligible threshold effects might be present.
It is obvious that this
limit $sin \theta=1$ is quite predictive.
For a vanishing cosmological constant (i.e.
$C=1$), the soft terms are in the ratio $m_i:M_a:A=1:{\sqrt 3}:{-\sqrt 3}$
up to small threshold effect corrections (and neglecting phases). This will
result in definite patterns for the low-energy particle
spectra\cite{Barbieri,BIM}.

\subsection{Computing soft terms in specific Superstring models}

In order to obtain more concrete expressions for the soft terms one has to
compute the functions $N_i(T,T^*)$, $\omega _{ijk}(T,T^*)$ and $f_a(S,T)$. In
order
to evaluate these functions one needs a minimum of information about the
K\"ahler potential $K$, the structure of Yukawa couplings $Y_{ijk}(T)$ and the
one-loop threshold corrections $f_a(T)$. This type of information is only
known for some classes of four-dimensional Superstrings which deserve special
attention.
An extended study of two large classes of models,
the large--$T$ limit of Calabi-Yau compactifications and orbifold
compactifications, can be found in ref.\cite{BIM}. The one--loop (Superstring)
corrections to the K\"ahler potential in orbifolds, which mix $S$ and
$T$ fields and as a consequence modify the tree--level soft terms, are also
analyzed. The general pattern of SUSY--spectra found (assuming vanishing
\footnote{The impact of a non--vanishing tree--level
cosmological constant on SUSY--spectra was studied in ref.\cite{Park}.}
cosmological constant)
is very characteristic.

\subsection{Computing soft terms in the case with several moduli ($T_m)$}

In the case with several moduli ($T_m$) the situation is more cumbersome
and one is forced to define new goldstino angles. This was first done in
section 8 of ref.\cite{BIM} in a different context (extra matter fields).
Following this line, eq.(\ref{F28}) is modified to\cite{Unpub,Japoneses}
\begin{eqnarray}
({G}^S_S)^{1/2}\ F^S\ &=&\ {\sqrt 3}C m_{3/2}\ e^{i\alpha _S}sin\theta
\nonumber\\
({G}^{T_m}_{T_m})^{1/2}\ F^{T_m}\ &=&\ {\sqrt 3}C m_{3/2}\
e^{i\alpha _{T_m}}cos\theta \Delta_m
\label{F95}
\end{eqnarray}
For instead, for $m=4$ (this is e.g. the case of some $Z_N$ and $Z_N \times
Z_M$
orbifolds with three
diagonal (1,1) moduli and one (2,1) moduli), three new goldstino angles are
necessary: $\Delta_1=cos\theta_1, \Delta_2=sin\theta_1 cos\theta_2,
\Delta_3=sin\theta_1 sin\theta_2 cos\theta_3, \Delta_4=sin\theta_1
sin\theta_2 sin\theta_3$.
Now, with these definitions, the computation of the soft terms gives
\begin{eqnarray}
M_a &=& {\sqrt 3}m_{3/2}C({{k_a ReS}\over {Ref_a}}e^{-i\alpha _S}sin\theta
+ cos\theta \sum_m e^{-i\alpha_{T_m}} {{b_a'^m(T_m+T^*_m)
{\hat {G_2}}(T_m,T^*_m)}\over{32\pi ^3 Ref_a}} \Delta_m
\nonumber\\
m_i^2 &=& m_{3/2}^2C^2(1\ +3 cos^2\theta\ \sum_m  n_i^m \Delta_m^2)\
+\ 2m_{3/2}^2(C^2-1)
\nonumber\\
A_{ijk} &=& -{\sqrt 3} C m_{3/2}( e^{-i\alpha _S}sin\theta\ +\
cos\theta \sum_m e^{-i\alpha _{T_m}} \Delta_m \omega _{ijk}(T_m)\ )\
\nonumber\\
\omega _{ijk}(T_m) &=& 1+n_i^m+n_j^m+n_k^m\ -(T_m+T^*_m)
{{Y_{ijk}^{T_m}}\over {Y_{ijk}}}
\label{F97}
\end{eqnarray}
where $n_i$ are the modular weights of the matter fields. A more complete
analysis including the B--term, one--loop (Superstring) corrections,
phenomenological consequences, and a comparison with the overall modulus
($T$) case can be found in ref.\cite{BIMS}.

\section{Gauge coupling constants unification in Superstring models}

The experimental (LEP) results agree with the joining of the three gauge
coupling constants of the MSSM at a single unification scale
($\sim 10^{16} GeV$).
In Superstring theory the coupling constants are unified even in the absence
of a GUT\footnote{Thus GUT gauge groups, as e.g. SU(5) or SO(10), are not
mandatory in order to have unification in the context of Superstring theory.},
since $ReS=1/(k_ag_a^2)$. However, this unification takes place at a scale
($\sim 10^{17} GeV$), i.e. one order or magnitude of difference. Perhaps,
to worry about a discrepancy $\sim 10$ when we are working with energy scales
$\sim 10^{16}$ is nonsense. Anyway,
three possible mechanisms have been proposed in order to explain this
discrepancy in the context of the MSSM from Superstrings.

First, the normalization factor of the hypercharge ($k_1$) which is fixed
in usual GUTs (e.g. $k_1=5/3$ in SU(5)), is a model--dependent number
in Superstring models. Playing around with this number we could obtain the
correct Weinberg angle and $\alpha_S$ (taking into account the experimental
uncertainties) at low--energy. This mechanis was first proposed in
ref.\cite{CM} and analyzed in detail in ref.\cite{Ibanez}. Second, large
Superstring threshold effects could produce different values of the
coupling constants at the Superstring unification scale but the same ones
\cite{Ross,IL} at a scale $\sim 10^{16} GeV$. Finally, one can consider further
charged particles\cite{Ellis} apart from those of the MSSM. This will require
the
introduction of some intermediate mass scale(s).

In the context of Superstring GUTs the only possibility would be to have
an intermediate scale $\sim 10^{16} GeV$ at which a GUT symmetry
like SU(5) of SO(10) is at work.

\end{document}